\input epsf
\documentstyle[preprint,eqsecnum,aps]{revtex}

\begin{document}

\count255=\time\divide\count255 by 60 \xdef\hourmin{\number\count255}
  \multiply\count255 by-60\advance\count255 by\time
  \xdef\hourmin{\hourmin:\ifnum\count255<10 0\fi\the\count255}

\draft
\preprint{\vbox{\hbox{WM-99-107}\hbox{JLAB-THY-99-11}
}}

\title{A Hexagonal Theory of Flavor}

\author{Christopher D. Carone$^{\dagger}$ and Richard
F. Lebed$^\ddagger$}

\vskip 0.1in

\address{$^\dagger$Nuclear and Particle Theory Group, Department of
Physics, College of William and Mary, Williamsburg, VA 23187-8795\\
$^\ddagger$Jefferson Lab, 12000 Jefferson Avenue, Newport News, VA
23606}

\vskip .1in
\date{May, 1999}
\vskip .1in

\maketitle
\tightenlines
\thispagestyle{empty}

\begin{abstract}
We construct a supersymmetric theory of flavor based on the discrete
gauge group $(D_6)^2$, where $D_6$ describes the symmetry of a regular
hexagon under proper rotations in three dimensions.  The
representation structure of the group allows one to distinguish the
third from the lighter two generations of matter fields, so that in
the symmetry limit only the top quark Yukawa coupling is allowed and
scalar superpartners of the first two generations are degenerate.
Light fermion Yukawa couplings arise from a sequential breaking of the
flavor symmetry, and supersymmetric flavor-changing processes remain
adequately suppressed.  We contrast our model with others based on
non-Abelian discrete gauge symmetries described in the literature, and
discuss the challenges in constructing more minimal flavor models
based on this approach.
\end{abstract}


\newpage
\setcounter{page}{1}

\section{Introduction}

        Perhaps the most puzzling issue in the physics of elementary
particles is the origin of the fermion mass hierarchy. The top quark
mass is observed to be $350,000$ times larger than that of the
electron, although both originate at the scale at which electroweak
symmetry is spontaneously broken.  Other quark and lepton mass
eigenvalues, as well as the off-diagonal elements of the
Cabibbo-Kobayashi-Maskawa (CKM) matrix exhibit hierarchies that are no
less mysterious.  While many models have been proposed to explain the
flavor structure of the standard model, a theory that is economical
and compelling has yet to be found.

        The naturalness of small couplings, in this case the Yukawa
couplings of the light fermions, can be understood in a
field-theoretic context if a symmetry of the Lagrangian is restored
when these couplings are set to zero.  Thus, one approach to the
flavor problem is to propose new symmetries acting horizontally
across standard model generations, that forbid all but the top quark
Yukawa coupling in the flavor symmetric limit.  If the flavor symmetry
group $G_f$ is spontaneously broken by a set of ``flavon'' fields
$\Phi$, then light fermion masses arise generically via higher-dimension 
operators involving the $\Phi$ fields, together with the
matter and Higgs fields of the standard model.  These operators may
have a renormalizable origin, as in the Froggatt-Nielsen
mechanism~\cite{fn}, or may simply be present at the Planck scale.  A
hierarchy in the light fermion Yukawa couplings may be explained if
$G_f$ is broken sequentially through a series of smaller and smaller
subgroups, at successively lower energy scales. In this paper we
consider a model that explains the quark and lepton mass hierarchies
in precisely this way.  In addition, we work within the context of
supersymmetry so that the hierarchy between the weak scale and the
flavor physics scale $M_F$ is stable under radiative corrections.

        One must now choose a type of symmetry, either global or
gauged, for building a model of flavor.  Global symmetries are thought
to be violated by quantum gravitational effects, and may not be
consistent as fundamental symmetries of nature~\cite{cgs}.  We
therefore focus on gauge symmetries in this paper.  A complication in
supersymmetry is that continuous gauge symmetries lead to $D$-term
interactions between the flavon fields and the scalar superpartners of
the minimal supersymmetric standard model (MSSM).  These couplings
lead to large off-diagonal squark masses, and unacceptable
flavor-changing neutral current (FCNC) effects~\cite{rf12}.
Fortunately, discrete gauge symmetries provide a viable alternative. A
discrete gauge symmetry can arise, for instance, when a continuous
gauge symmetry group is spontaneously broken to a discrete subgroup.
In this case, the discrete charges of fields in the low-energy
effective theory must satisfy constraints that follow from the anomaly
freedom of the original gauge group~\cite{IR}.  More importantly,
however, discrete gauge symmetries can be defined independently of any
embedding~\cite{BD}.  For example, such symmetries might arise
directly in string compactifications.  We construct a model based on a
discrete gauge symmetry satisfying the appropriate low-energy
constraints, independent of the origin of the symmetry.

        In addition, we restrict our attention to non-Abelian discrete
gauge groups, since these groups can have two-dimensional irreducible
representations.  By placing the first two generations of quark
superfields into a doublet, for example, we achieve exact degeneracy
of the corresponding squarks in the flavor symmetry limit~\cite{DLK}.
Since $G_f$ is broken only by light fermion Yukawa couplings, a high
degree of degeneracy is preserved in the low-energy theory.  In this
way, one greatly suppresses the FCNC effects that one would expect
otherwise if the superparticle spectrum were generic.

        Despite the expansive literature on models of
flavor~\cite{S33,FKong,KS,FKep,otherN,otherA}, relatively few
successful supersymmetric models based on discrete non-Abelian gauged
flavor symmetries exist~\cite{S33,FKong,KS}.  It seems that most
models fall into two categories: (a) models with relatively large
discrete groups containing so many subgroups that they can easily
accommodate the observed fermion mass hierarchy via a sequential
breaking of the symmetry, and (b) models based on smaller discrete
symmetries that require fine tuning to compensate for the fact that
the flavor symmetry is a bit too restrictive.  The $(S_3)^3$
model~\cite{S33}, for example, is interesting in that it is based on
the smallest non-Abelian group $S_3$, and has a simple representation
structure.  However, the replication of $S_3$ factors yields a
discrete group with $(3!)^3 = 216$ elements and numerous subgroups,
and so it is not surprising that one can find a breaking pattern that
accommodates viable Yukawa textures.  On the other hand the
$Q_{12}\times U(1)_H$ model~\cite{FKong} has a smaller discrete group
(one with $120$ elements, when one takes into account that only a
$Z_5$ subgroup of the U(1) is relevant), but has quark mass textures
that would not be viable without tuning of order one operator
coefficients.  Our goal is to find a model that lives somewhere
between the two extremes defined above, and believe that the model we
present in this paper is a step in the right direction.  We hope that
it will lead to interest in finding an optimal theory of flavor based
on a non-Abelian discrete gauge group.

        Our model is based on the group $D_6 \times D_6$, where $D_6$
describes the symmetry of a regular hexagon under proper rotations in
three dimensions.  In Sec.~\ref{d6}, we discuss this symmetry group in
more detail, and explain why it is promising for model building. The
model itself is presented in Sec.~\ref{model}.  Left- and right-handed
matter fields transform under separate $D_6$ factors, and the three
generations of a given field have a simple doublet-plus-singlet
representation structure, with quarks and leptons distinguished by
their transformation properties under a parity subgroup.  In
Sec.~\ref{phenom} we consider bounds on the model from flavor-changing
processes, and in the final section we summarize our conclusions.
Fundamental properties of the theory of discrete groups are summarized 
in Appendix~\ref{grpthy}, and details of combining $D_6$ representations 
appear in Appendix~\ref{clebsch}.

\section{The Group $D_6$} \label{d6}

	In order to obtain the degree of degeneracy among the scalar
superpartners necessary to suppress FCNC's at a phenomenologically
acceptable level, we require that the light two generations of a given
MSSM matter field appear in at least a doublet. All Abelian groups can
be shown to possess only one-dimensional irreducible representations,
and therefore the desired horizontal symmetry group $G_f$ must be
non-Abelian.  Although it is tempting to explain the three-family
replication by placing all quarks in triplet representations of $G_f$,
it is also true that the top quark appears to be distinguished from
the others by its large mass.  Indeed, the top quark is the only one
in the standard electroweak theory with an $O(1)$ Yukawa coupling,
suggesting that it may have different transformation properties than
fields of the first two generations.  Thus, it is natural to assign
the top quark to a (possibly nontrivial) one-dimensional
representation of $G_f$. If one works only with triplet quark
representations as in Ref.~\cite{KS}, one must explain with dynamics
rather than symmetry why $m_t$ is so large.

\begin{figure}
  \begin{centering}
  \def\epsfsize#1#2{1.50#2}
  \hfil\epsfbox{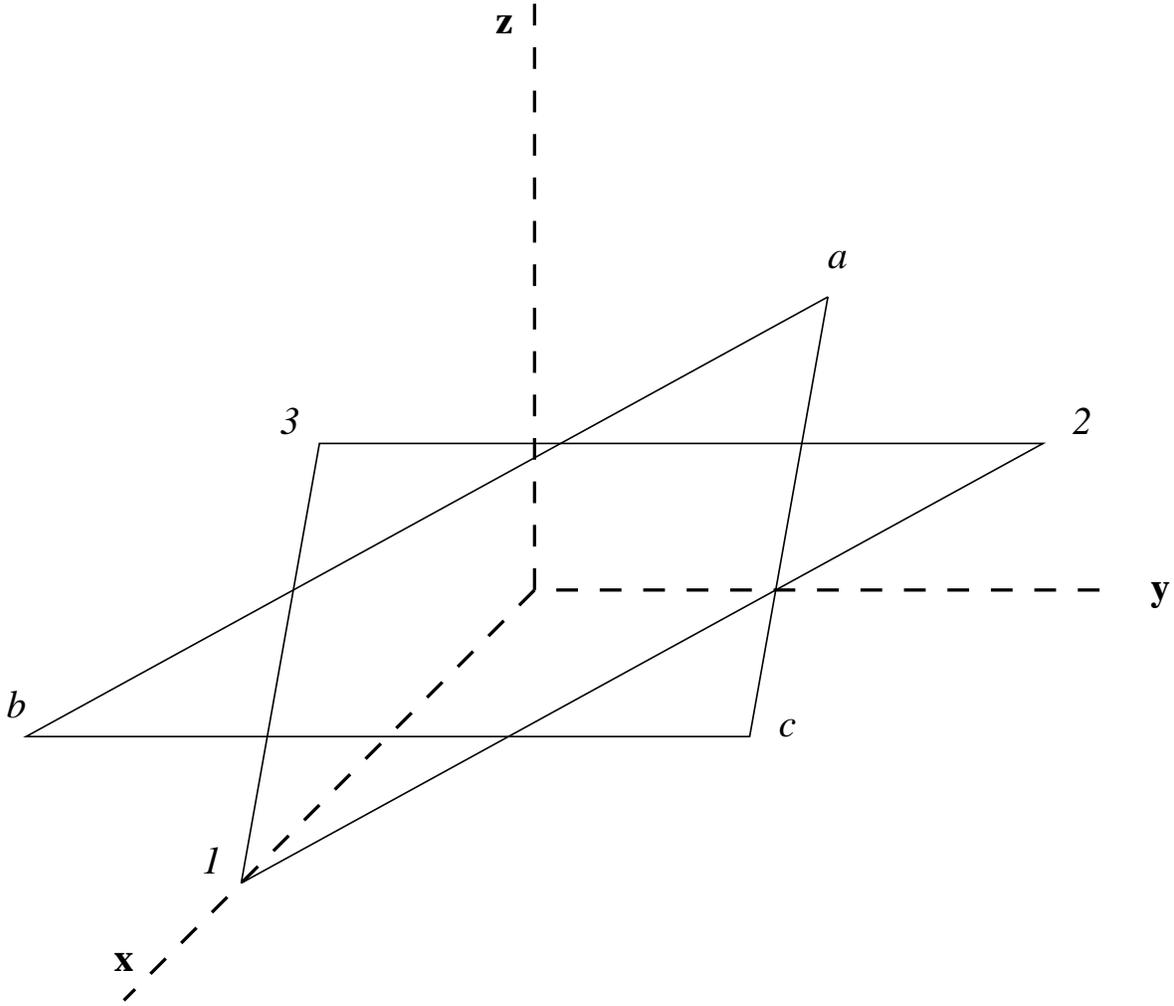}\hfill

\caption{Graphical representation of the groups $D_3$ and $D_6$.  In
terms of permutation cycle notation ({\it e.g.}, (132)(45) means $1
\to 3$, $3 \to 2$, $2 \to 1$, $4 \longleftrightarrow 5$), $D_3$ is
generated by $C_3$, a $2\pi/3$ rotation about ${\bf \hat z}$, which is
equivalently (123), and $C_x$, a $\pi$ rotation about ${\bf \hat x}$,
which is (1)(23).  Similar statements apply to the triangle $abc$,
which is disjoint from 123 as far as $D_3$ is concerned.  $D_6$
adjoins the rotation $C_2$, a $\pi$ rotation about ${\bf \hat z}$,
which is (1$a$)(2$b$)(3$c$).  Alternately, $D_6$ is generated by the
$\pi/6$ rotation $C_6 = C_2 C_3^{-1}$ = (1\,$c$\,2\,$a$\,3\,$b$) and
$C_x = (1)(23)(a)(bc)$.\label{fig1}}

  \end{centering}
\end{figure}

	The smallest non-Abelian group is $S_3$, the group of
permutations among three objects.  It has 3! = 6 elements, and
possesses one doublet and two singlet representations, thus satisfying
Eq.~(\ref{repcount}).  $S_3$ is isomorphic to the dihedral group
$D_3$, which is the group of proper rotations in three dimensions that
leave an equilateral triangle invariant.  To picture this
geometrically (Fig.~\ref{fig1}), let the triangle lie in the
$xy$-plane with its centroid at the origin, so that $\hat {\bf z}$ is
the normal vector.  Then $D_3$ is generated by a rotation ($C_3$) of
$2\pi/3$ radians about $\hat {\bf z}$ and a rotation ($C_x$)
of $\pi$ radians about $\hat {\bf x}$ that turns the triangle
upside-down.  Note that $C_3 C_x \ne C_x C_3$, so that the group is
non-Abelian.  The three irreducible representations in this picture
are ${\bf 1_S}$, the trivial singlet left invariant by all actions of
the group; ${\bf 1_A}$, a singlet invariant under $C_3$ but odd under
$C_x$ (such as the vector $\hat {\bf z}$); and a doublet ${\bf 2}$,
one realization of which is the pair ($\hat {\bf x}$, $\hat {\bf y}$).

        The simplicity of this structure was used to great effect in
the $(S_3)^3$ model~\cite{S33}; there, the three generations of
left-handed quarks $Q$, the right-handed up-type quarks $U$, and the
right-handed down-type quarks $D$ were each assigned to ${\bf 2} +
{\bf 1_A}$ representations of different $S_3$ factor groups, while
remaining trivial (transforming as ${\bf 1_S}$) under the others.  The
left-handed leptons $L$ and their right-handed partners $E$ were
assigned to ${\bf 2} + {\bf 1_A}$ representations of $S_3^D$ and
$S_3^Q$, respectively.  As one sees from the previous paragraph, the
combination ${\bf 2} + {\bf 1_A}$ has the natural geometric
interpretation of forming a complete vector in three-dimensional
space, if one pictures $D_3$ embedded in the group O(3) of all
rotations in three dimensions.  This fact is useful in understanding
anomaly cancellation conditions, as is discussed in
Section~\ref{model}.

        The $(S_3)^3$ model successfully explains the light squark
degeneracy and the unique largeness of $m_t$, and also accommodates a
phenomenologically viable texture of Yukawa matrices through
sequential spontaneous symmetry breaking in which flavons,
transforming nontrivially under $G_f$, obtain vacuum expectation
values (VEV's).  For example, the flavon VEV that gives rise to the
bottom Yukawa coupling appears at a higher scale than the one that
gives rise to that of the strange quark, explaining why $m_s < m_b$.
The threefold replication of $S_3$ factors is necessary to allow for a
sufficient number of stages of symmetry breaking to accommodate all
the distinct Yukawa couplings.  At each stage of symmetry breaking
there remains some smaller set of symmetries still obeyed by the
Lagrangian of the theory, which is specified by some subgroup of the
original group $G_f$.  Because of its small size, a single $S_3$
factor is insufficient since $S_3$ contains few nontrivial subgroups
and therefore allows few distinct stages of symmetry breaking.  While
it is clearly possible to choose arbitrarily large groups that are
suitable for our purposes, such choices violate the desire for
minimality in particle physics.  Therefore, our goal is to find a
group large enough to accommodate all the properties in which we are
interested but small enough to be compelling for study as a
potentially real symmetry of nature.  We seek a model satisfying the
following criteria:
\begin{itemize}
\item Degeneracy of first two generations of squarks in the symmetry
limit: a ${\bf 2} + {\bf 1}$ representation structure for the up-type
quark superfields, and either a ${\bf 2} + {\bf 1}$ or ${\bf 3}$ for
the down-type quark and lepton superfields;

\item Invariance of only the top quark Yukawa coupling under $G_f$;

\item A simple pattern of anomaly cancellation;

\item Enough subgroups to accommodate a viable Yukawa texture via
sequential symmetry breaking;

\item Absence of unnatural fine tuning of parameters;

\item As small a group as possible, with minimal replication of factor
groups.

\end{itemize}
        In order to maintain the attractive features of the group
$D_3$, we consider the 12-element dihedral group $D_6$, which is the
smallest group that nontrivially contains $D_3$ as a subgroup.  $D_6$
is the symmetry group of a regular hexagon under proper rotations in
three dimensions, and indeed, is isomorphic to $D_3\times Z_2$, as may
be seen in Fig.~\ref{fig1}.  By adjoining $C_2$, the odd element of
the $Z_2$ group, which is a rotation by $\pi$ in the $xy$-plane, one
generates the complete $D_6$: The sixfold rotation ($\pi/3$ radians)
necessary for the hexagonal symmetry is $C_6 = C_2 C_3^{-1}$.

        $D_6$ possesses four singlet and two doublet inequivalent
irreducible representations.  Using the isomorphism $D_6 \cong D_3
\times Z_2$, these may be labeled {\bf R}$^P$, where {\bf R} is a
representation of $D_3$ and $P$ is a parity indicating the action of
$C_2$ on the elements of {\bf R}.  For example, the original
representations of $D_3$ transforming like $\hat {\bf z}$ and ($\hat
{\bf x}, \hat {\bf y}$) are identified with the $D_6$ representations
${\bf 1_A^+}$ and ${\bf 2^-}$, respectively.  The other
representations of $D_6$ are ${\bf 1_S^+}$ (the trivial singlet),
${\bf 1_S^-}$, ${\bf 1_A^-}$, and ${\bf 2^+}$.  Explicit matrix
representations and the rules for combining them are relegated to
Appendix~\ref{clebsch}.

        A real advantage of the dihedral groups $D_n$ is that, in each
doublet representation, the rotation matrix $C_x$ in a particular
basis (corresponding geometrically to a rotation of $\pi$ radians
about the $x$-axis) has the form
\begin{equation} \label{Z2gen}
C_x = \left(
\begin{array}{cc}
1 & 0 \\ 0 & -1
\end{array}
\right) ,
\end{equation}
which means that the first component of the doublet in this basis is
left invariant under the $Z_2$ subgroup of $D_n$ generated by the
twofold rotation $C_x$.  In particular, breaking $D_n \to Z_2$ permits
a VEV in the first component of the doublet while forbidding one in
the second.  This feature goes a long way toward building a
hierarchical Yukawa texture.

        There exists another 12-element group that generalizes $D_3$,
the ``dicyclic'' group $Q_6$.  Just as the full rotation group SU(2)
in quantum mechanics forms a double-covering of the classical rotation
group O(3), $Q_6$ may be thought of as the quantum-mechanical
generalization of $D_3$: In either SU(2) or $Q_6$, a $2\pi$ rotation
is not the identity operator for all representations, but rather there
exist ``spinorial'' representations for which such a rotation produces
a phase of $-1$.  In particular, one of the doublet representations of
$Q_6$ is spinorial.  Unfortunately, one can show that there is no
basis in which only one of the components of this doublet receives a
VEV, leaving an unbroken subgroup.  Therefore, the group $D_6$ is more
convenient than $Q_6$ for building a viable Yukawa texture.

        Finally, we have found that including only two factors of
$D_6$ is enough to satisfy all the listed criteria; in particular, we
have assigned $Q$ and $L$ to transform nontrivially under a single
$D_6$, called $D_6^L$ to denote chirality, and $U$, $D$, and $E$
under a corresponding $D_6^R$.  With $12^2 = 144$ elements, $D_6^L
\times D_6^R$ is the smallest group yet discussed in the literature
that satisfies our itemized criteria.  While we cannot rule out that
some single, smaller discrete group might be found that works equally
well in meeting our requirements, the simple, left-right symmetric
form of the $D_6^L \times D_6^R$ model presented in the next section
is intriguing, and suggests that this and similar group structures may
be worthy of further consideration.

\section{The $D_6^L \times D_6^R$ Model} \label{model}

\subsection{Field Content}

        We propose the flavor symmetry group $D_6^L \times D_6^R$,
where the three generations of matter fields transform as ${\bf 2}
+ {\bf 1}$-dimensional combinations of $D_6$.  Specifically,
\begin{eqnarray}
Q & \sim & ( [ {\bf 2^-}, {\bf 1_S^+} ], [ {\bf 1_A^+}, {\bf 1_S^+} ]
) , \nonumber \\
U & \sim & ( [ {\bf 1_S^+}, {\bf 2^-} ], [ {\bf 1_S^+}, {\bf 1_A^+}
] ), \nonumber \\
D & \sim & ( [ {\bf 1_S^+}, {\bf 2^-} ], [ {\bf 1_S^+}, {\bf 1_A^+}
] ), \nonumber \\
L & \sim & ( [ {\bf 2^+}, {\bf 1_S^+} ], [ {\bf 1_A^-}, {\bf 1_S^+} ]
) , \nonumber \\
E & \sim & ( [ {\bf 1_S^+}, {\bf 2^+} ], [ {\bf 1_S^+}, {\bf 1_A^-}
] ) ,
\label{eq:matter}
\end{eqnarray}
where the two entries inside square brackets distinguish the $D_6^L$
and $D_6^R$ transformation properties of each field, and the two sets
of square brackets themselves refer to the transformation properties
of the first two and third generation matter fields, in that order.
Note that the left-handed fields $Q$ and $L$ transform nontrivially
only under $D_6^L$, while the right-handed fields $U$, $D$, and
$E$ transform nontrivially only under $D_6^R$.  Note also that quark
and lepton fields are distinguished by their $D_6$ parity properties:
Quarks transform as ${\bf 2^-} + {\bf 1_A^+}$, while leptons
transform as ${\bf 2^+} + {\bf 1_A^-}$.

        In contrast, the MSSM Higgs fields must transform under both
$D_6$ symmetries:
\begin{eqnarray}
H_U & \sim & ( {\bf 1_A^+}, {\bf 1_A^+} ) , \nonumber \\
H_D & \sim & ( {\bf 1_A^-}, {\bf 1_A^+} ) .
\label{eq:higgs}
\end{eqnarray}
The differing parity assignments assure that the top quark Yukawa
coupling is invariant under the flavor symmetry, while that of the
bottom quark is forbidden in the symmetry limit.

	It is worth mentioning that the assignment of, for example,
$L$ and $D$ to different $D_6$ factors implies that this model
cannot be embedded easily into a unified field theory containing SU(5)
as a subgroup.  However, our model is perfectly consistent with gauge
unification in a string theoretic context, which some may argue is
preferable, given the absence of the usual doublet-triplet and proton
decay problems associated with the existence of color-triplet Higgs
fields.

\subsection{Anomaly Cancellation}

	To understand anomaly cancellation in this model, it is useful
to study first a simpler example. Consider a toy model based on the
flavor symmetry $G_f=S_3$.  Let us assign the three generations of
matter fields to ${\bf 2}+{\bf 1_A}$ representations, and assign the
two Higgs fields of the MSSM, $H_U$ and $H_D$, both to ${\bf 1_A}$'s.
We wish to show that $S_3$ is a consistent discrete gauge symmetry in
this low-energy theory.  To do so, imagine adding two additional Higgs
fields, $H_U^{(2)}$ and $H_D^{(2)}$, which are doublets under $S_3$.
Now one can embed all the fields into {\bf 3}-dimensional
representations of O(3), since a {\bf 3} decomposes into ${\bf 2}+{\bf
1_A}$.  It is clear in this case that the O(3)$\times$SU(2)$^2$ and
O(3)$\times$SU(3)$^2$ anomalies vanish trivially, since the O(3)
generators are antisymmetric, and hence traceless.  It was
demonstrated by Banks and Dine \cite{BD} that the only anomaly
cancellation conditions needed for the consistency of the low-energy
effective theory are those derived by Ib\'{a}\~{n}ez and
Ross~\cite{IR} that are linear in the embedding group (in this case
O(3)), and that involve only the non-Abelian low-energy continuous
gauge groups.  Furthermore, they point out that this is equivalent to
the requirement that any effective interaction ${\cal O}_{\rm eff}$
generated by instantons of these continuous gauge groups remains
invariant under the action of the discrete group.  Since this latter
criterion does not depend on any particular embedding of the discrete
group into a continuous one at higher energies, it is the actual
constraint we demand on the low-energy effective theory.  Since we
have seen that the linear O(3) anomalies vanish, we conclude that
$S_3$ is not anomalous in the extended version of the toy model.  Now
notice that the Higgs fields in the extended theory form a vector-like
pair, so that one can give a large mass to the unwanted doublet
components, thus decoupling them from the low-energy theory.  (In the
O(3) embedding, this can be accomplished by breaking O(3) through
giving a large VEV to only the linear combination of {\bf 1}+{\bf 5}
coupling to the doublets.)  Since $S_3$ remains unbroken when
integrating out the unwanted states, the set of operators generated
from ${\cal O}_{\rm eff}$ involving just the light fields, ${\cal
O}^\prime_{\rm eff}$, also remains invariant under $S_3$.  Thus, one
concludes that desired low-energy theory is also free of discrete
gauge anomalies.  This is the basis of the $S_3$ charge assignments in
the $(S_3)^3$ model described in Refs.~\cite{S33}.

        In our model, we maintain the multiplicity of {\bf 2} and
${\bf 1_A}$ representations described in the toy model above, for any
given $S_3$ factor.  However, the group $D_6$ is isomorphic to
$S_3\times Z_2$, and so one must also check that the $Z_2$ factor is
not anomalous.  To do so, one can embed $Z_2$ into a U(1) and check
that the U(1) charges satisfy the linear Ib\'{a}\~{n}ez-Ross condition
for the $Z_2 \times {\rm SU}(N)^2$ anomaly\cite{IR}:
\begin{equation}
\sum_i T_i (q_i) = 2 r \,\, , \,\,\,\,\,\,\, \mbox{integer r}
\label{eq:zanom}
\end{equation}
where $T_i$ is the SU($N$) invariant defined by ${\rm Tr} \, t^a t^b =
T \delta^{ab}$ and $q_i$ is the U(1) charge of the $i$th field.  For
example, the right-handed quark fields might be embedded into ${\bf 2
+ 1_{A}}$, where we assign U(1) charges $+1$ and $+2$ to the ${\bf 2}$
and ${\bf 1_A}$, respectively. Then Eq.~(\ref{eq:zanom}) tells us
\[
\frac{1}{2}(2 \cdot 1 + 2) = 2 r
\]
which is satisfied for $r=1$.  Thus, the $D_6$ representation ${\bf
2^- + 1_A^+}$ is anomaly free.  It is straightforward to verify that
Eq.~(\ref{eq:zanom}) is satisfied for SU(2) and SU(3) given the $Z_2$
charge assignments presented in (\ref{eq:matter}) and
(\ref{eq:higgs}).  Note that the anomaly from the $Z_2$ subgroup
of $D_6^L$ cancels between the third-generation $L$ field and $H_D$.

\subsection{Yukawa Textures}

	To derive the Yukawa textures, one requires the rules for
combining $D_6$ representations; these are presented in
Appendix~\ref{clebsch}.  Treating the Yukawa matrices $Y$ as flavor
spurions, the superpotential terms $Q Y_U H_U U$, $Q Y_D H_D D$, and
$L Y_L H_D E$ must be trivial singlets under $G_f$.  Thus, the flavon
fields contributing to the Yukawa matrices must transform as
\begin{eqnarray} \label{Yuk}
Y_U & \sim & \left( \begin{array}{cc} [ {\bf \tilde 2^-}, {\bf \tilde
2^-}] & [ {\bf \tilde 2^-}, {\bf 1_S^+} ] \\ {[} {\bf 1_S^+}, {\bf
\tilde 2^-} ] & [ {\bf 1_S^+}, {\bf 1_S^+} ]
\end{array} \right) ,
\nonumber \\
Y_D & \sim & \left( \begin{array}{cc} [ {\bf \tilde 2^+}, {\bf \tilde
2^-}] & [ {\bf \tilde 2^+}, {\bf 1_S^+} ] \\ {[} {\bf 1_S^-}, {\bf
\tilde 2^-} ] & [ {\bf 1_S^-}, {\bf 1_S^+} ] \end{array} \right) ,
\nonumber \\
Y_L & \sim & \left( \begin{array}{cc} [ {\bf \tilde 2^-}, {\bf \tilde
2^+}] & [ {\bf \tilde 2^-}, {\bf 1_S^-} ] \\ {[} {\bf 1_S^+}, {\bf
\tilde 2^+} ] & [ {\bf 1_S^+}, {\bf 1_S^-} ] \end{array} \right) .
\end{eqnarray}
Given a doublet ${\bf 2} = (a,b)$, the conjugate doublet ${\bf \tilde
2} = (-b, a)$ arises through the product of a ${\bf 2}$ with a ${\bf
1_A}$, as described in Appendix~\ref{clebsch}.

        For comparison, the squark and slepton mass matrices, which
appear in the Lagrangian as terms of the form $\tilde \phi^\dagger
m_{\tilde \phi}^2 \tilde \phi$, where $\tilde \phi$ is the scalar
component of the superfield $\phi = Q, U, D, L,$ or $E$, transform as
\begin{eqnarray} \label{sqk}
m_{\tilde Q}^2, m_{\tilde L}^2 & \sim & \left( \begin{array}{c|c} [
{\bf 2^+}, {\bf 1_S^+} ] \oplus [ {\bf 1_A^+}, {\bf 1_S^+} ] \oplus [
{\bf 1_S^+}, {\bf 1_S^+}] & [ {\bf \tilde 2^-}, {\bf 1_S^+} ] \\
\hline {[} {\bf \tilde 2^-}, {\bf 1_S^+} ] & [ {\bf 1_S^+}, {\bf
1_S^+} ] \end{array} \right) , \nonumber \\
m_{\tilde U}^2, m_{\tilde D}^2, m_{\tilde E}^2 & \sim & \left(
\begin{array}{c|c} [ {\bf 1_S^+}, {\bf 2^+} ] \oplus [ {\bf 1_S^+},
{\bf 1_A^+} ] \oplus [ {\bf 1_S^+}, {\bf 1_S^+} ] & [ {\bf 1_S^+},
{\bf \tilde 2^-} ] \\ \hline {[} {\bf 1_S^+}, {\bf \tilde 2^-} ] & [
{\bf 1_S^+}, {\bf 1_S^+} ]
\end{array} \right) ,
\end{eqnarray}
where the internal lines divide the first two generations from the
third.

        One sees from Eq.~(\ref{Yuk}) that of 35 possible nontrivial
flavon structures, only 11 appear in the Yukawa matrices.  Two more
(see Eq.(\ref{sqk})) appear in the squark mass matrices, and the
remaining 22 cannot appear at tree level in any of the matrices.
Indeed, we claim that a phenomenologically viable structure may be
obtained through the introduction of only 5 types of flavon fields,
\begin{eqnarray}
\sigma_b & \sim & [ {\bf 1_S^-}, {\bf 1_S^+} ], \nonumber \\
\sigma_\tau & \sim & [ {\bf 1_S^+}, {\bf 1_S^-} ], \nonumber \\
\phi & \sim & [ {\bf \tilde 2^-}, {\bf 1_S^+} ] , \nonumber \\
\psi & \sim & [ {\bf 1_S^+}, {\bf \tilde 2^-} ] , \nonumber \\
\Sigma & \sim & [ {\bf \tilde 2^+}, {\bf \tilde 2^+} ] .
\end{eqnarray}
All of the other flavon fields necessary to achieve a satisfactory
accounting of phenomenology arise as products of these, at higher
order in the effective Lagrangian.

        The pattern of symmetry breaking is likewise chosen to be
rather minimal; it consists of only three steps, and is summarized in
Table~\ref{sym}.  The level of symmetry breaking is described in
powers of a dimensionless parameter $\lambda$, and describes the size
of the corresponding VEV arising at that point relative to some high
energy scale $M_F$ at which perfect flavor symmetry is restored.  In
phenomenological terms, we choose $\lambda$ to be approximately the
size of the Cabibbo angle, $\lambda \approx 0.22$, and assign powers
of $\lambda$ consistent with fermion masses renormalized at a high
($\agt 10^{16}$ GeV) scale\cite{S33}.

        The full symmetry $D_6^L \times D_6^R$ is broken at the scale
$\lambda^2 M_F$ down to $Z_2^L \times Z_2^R$.  Here, the $Z_2$ factors
refer to the subgroup of $D_3 \subset D_6$ described in and after
Eq.~(\ref{Z2gen}), which allow one to give VEV's to the first
components of ${\bf 2}$'s, or equivalently the second components of
${\bf \tilde 2}$'s, while disallowing VEV's in the other components;
it is not to be confused with the $Z_2$ parity subgroup of $D_6$,
which is already broken after the first step.  Since this step breaks
both $D_6$'s to $Z_2$'s, only ${\bf 1_S}$ singlets, the second 
components of ${\bf \tilde 2}$'s, and the (22) component of 
$[ {\bf \tilde 2}, {\bf \tilde 2} ]$ obtain $O(\lambda^2$) VEV's at this 
stage.  In the second stage, $Z_2^L$ is broken to $Z_1^L$ ({\it i.e.},
no 
symmetry), which gives VEV's of $O(\lambda^3)$ to the first component of 
$D_6^L$ doublet $\phi$ and the first component in the second column of 
$\Sigma$ (and also to the second components, but these are subleading). 
In 
the third stage, all remaining components previously protected receive
$O(\lambda^4)$ VEV's.  In summary, the textures of leading VEV's in
the flavon fields read
\begin{eqnarray}
&& \sigma_b, \sigma_\tau \sim \lambda^2, \nonumber \\
&&\phi^T_a \sim ( \lambda^3, \, \lambda^2 ), \ \ a=1,2, \nonumber \\
&& \psi \sim ( \lambda^4, \, \lambda^2 ), \nonumber \\
&& \Sigma \sim \left( \begin{array}{cc} \lambda^4 & \lambda^3 \\
\lambda^4 & \lambda^2 \end{array} \right) .
\end{eqnarray}
The presence of two $\phi$ doublets but only one $\psi$
doublet leads to a viable phenomenology, as described below;  The
generic label $\phi$ used henceforth refers to a potentially different
linear combination of $\phi_1$ and $\phi_2$ each time $\phi$ appears.  
With these assignments, it is a straightforward matter to find the 
leading contributions to the Yukawa and scalar mass-squared matrices 
with transformation properties given by Eqs.~(\ref{Yuk}) and
(\ref{sqk}), 
and then to perform the tensor products as explained in 
Appendix~\ref{clebsch} to obtain the textures for these matrices:
\begin{eqnarray}
Y_U & \sim & \left( \begin{array}{c|c} \phi \psi & \phi \\ \hline \psi
& 1 \end{array} \right) \to \left( \begin{array}{ccc} \lambda^7 &
\lambda^5 & \lambda^3 \\ \lambda^6 & \lambda^4 & \lambda^2 \\
\lambda^4 & \lambda^2 & 1 \end{array} \right) , \nonumber \\
Y_D & \sim & \left( \begin{array}{c|c} \sigma_\tau \Sigma + \psi
\Sigma & \sigma_b \phi + \phi \phi + \Sigma \Sigma \\ \hline \sigma_b
\psi & \sigma_b \end{array} \right) \to \left( \begin{array}{ccc}
\lambda^6 & \lambda^5 & \lambda^5 \\ \lambda^6 & \lambda^4 & \lambda^4
\\ \lambda^6 & \lambda^4 & \lambda^2 \end{array} \right) , \nonumber
\\
Y_L & \sim & \left( \begin{array}{c|c} \sigma_b \Sigma + \phi \Sigma &
\sigma_\tau \phi \\ \hline \sigma_\tau \psi + \psi \psi + \Sigma
\Sigma & \sigma_\tau \end{array} \right) \to \left(
\begin{array}{ccc} \lambda^6 & \lambda^5 & \lambda^5 \\ \lambda^6 &
\lambda^4 & \lambda^4 \\ \lambda^6 & \lambda^4 & \lambda^2 \end{array}
\right), \nonumber \\
m_{\tilde Q}^2, m_{\tilde L}^2 & \sim & \left( \begin{array}{c|c}
\begin{array}{c} \{ \sigma_b \phi + \phi \phi  + \Sigma \Sigma \} \\
\oplus \, \{ \Sigma \Sigma + \phi \phi \} \oplus \, 1 \end{array} &
\phi \\ \hline \phi^T & 1 \end{array} \right) \to M^2 \left(
\begin{array}{ccc} \lambda^0 + \lambda^4 & \lambda^5 & \lambda^3 \\
\lambda^5 & \lambda^0 + \lambda^4 & \lambda^2 \\ \lambda^3 & \lambda^2
& \lambda^0
\end{array} \right) , \nonumber \\
m_{\tilde U}^2, m_{\tilde D}^2, m_{\tilde E}^2 & \sim & \left(
\begin{array}{c|c} \begin{array}{c} \{ \sigma_\tau \psi
+ \psi \psi + \Sigma \Sigma \} \\ \oplus \, \{ \psi \psi + \Sigma
\Sigma \} \oplus \, 1 \end{array} & \psi^T \\ \hline \psi & 1
\end{array} \right) \to M^2 \left( \begin{array}{ccc} \lambda^0 +
\lambda^4 & \lambda^6 & \lambda^4 \\ \lambda^6 & \lambda^0 + \lambda^4
& \lambda^2 \\ \lambda^4 & \lambda^2 & \lambda^0
\end{array} \right) , \nonumber \\
\end{eqnarray}
where $M$ is the squark/slepton mass scale, assumed to be a few
hundred GeV.  The $O(\lambda^0)$ entries in the first two generations
for each superfield are identical as a consequence of the flavor
symmetry, while the $O(\lambda^0)$ entry in the third generation may
be different.

        In writing these matrices we have neglected all signs in the
exact Clebsch-Gordan couplings, which are irrelevant in an effective
theory unless there is a special correlation of the VEV's.  For our
purposes, this occurs only in the upper $2 \times 2$ block of $Y_U$,
where one sees that only one combination of flavons, $\phi \times
\psi$, occurs at leading order.  Observe that the $n \times n$ matrix
obtained from the product of an $n$-dimensional column vector with an
$n$-dimensional row vector has $n-1$ zero eigenvalues and only one
nonzero eigenvalue, the dot product of the two vectors.  In the
present case, a zero eigenvalue for the $2 \times 2$ matrix $\phi
\times \psi$ means that the up quark Yukawa coupling remains exactly
zero at this order.  Note that this would not be the case if more than
one
$\psi$ doublet were present.  On the other hand, at least two $\phi$
doublets are needed to guarantee that $Y_U$ does not have two zero
eigenvalues.  Indeed, this eigenvalue zero is lifted only
at third order in the flavons, where contributions appear from
\begin{equation}
\sigma_b \sigma_\tau \Sigma, \,\,\,\,\, \ \sigma_b \psi \Sigma,
\,\,\,\,\,
\ \sigma_\tau \phi \Sigma, \,\,\,\,\, \mbox{ and } \,\,\,\,\, \ \phi 
\psi \Sigma .
\end{equation}
The computation of these flavon products in all orderings
(Clebsch-Gordan coefficients are not in general associative) produces
the same texture in each case,
\begin{equation}
\left. \Delta Y_U \right|_{{\rm upper} \ 2 \times 2} = \left(
\begin{array}{cc} \lambda^8 & \lambda^7 \\ \lambda^8 & \lambda^6
\end{array} \right) ,
\end{equation}
thus lifting the zero eigenvalue and providing an up quark
Yukawa coupling of $O(\lambda^8)$.

\subsection{Non-Supersymmetric Phenomenology}

        Now let us consider how this assignment of fields and pattern
of VEV's leads to a set of quark masses and mixing angles that are
phenomenologically acceptable.

        First, observe that the Yukawa coupling $h_t$ giving rise to
the top quark mass is unique in surviving in the symmetry-conserving
$\lambda \to 0$ limit.  The largest parameters breaking the symmetry
arise at $O(\lambda^2)$, and give rise to $h_b$, $h_\tau$, and
$V_{cb}$.  In typical flavor models with $\tan \beta \approx 1$, $h_b$
and $h_\tau$ appear at $O(\lambda^3)$; their slightly larger size in
this model may easily be accommodated choosing $\tan \beta \approx 6$
or a smaller $\tan \beta$ combined with some adjustment of the
undetermined order unity coefficients.

        The next parameters of interest are $V_{ub}$, arising at
$O(\lambda^3)$ and $h_c$, $h_s$, and $h_\mu$ at $O(\lambda^4)$.  These
give rise to phenomenologically acceptable values for our choice of
$\tan \beta$.  Note at this point that the Yukawa matrices $Y_D$ and
$Y_L$, although arising from different flavon structures, have the
same basic texture; in order to accommodate the detailed differences
between the down-type quark and lepton textures, one must choose
coefficients such that $(Y_L)_{22}/(Y_D)_{22} \approx 3$ 
and $(Y_L)_{11}/(Y_D)_{11} \approx 1/3$.

        The entries giving rise to $V_{us}$, $(Y_U)_{12}$ or
$(Y_D)_{12}$, appear at $O(\lambda^5)$, and indeed give rise upon
diagonalization to a Cabibbo angle of $O(\lambda)$, since the (12)
entries are suppressed by one power of $\lambda$ compared to the (22)
entries.  The couplings $h_d$ and $h_e$ appear at $O(\lambda^6)$, and
have appropriate magnitudes, taking into account the factor 1/3
previously mentioned.  Finally, $h_u$ arises only at third order in
the flavon fields as described in the previous subsection, and appears
(as desired) at $O(\lambda^8)$.

\section{Phenomenology} \label{phenom}

        In the previous section, we showed that there is a three-step
symmetry breaking pattern for $D_6^L \times D_6^R$ that yields viable
Yukawa textures, given our choice of flavon fields.  It remains to be
addressed whether the scalar superpartner mass-squared matrices retain
a sufficient degeneracy after the flavor symmetry group is broken to
suppress supersymmetric FCNC effects adequately.  We consider this
question in the current section.  First, note that scalar degeneracy
is still approximate in the low energy theory, so one may treat
off-diagonal scalar masses in an insertion approximation.  This is
clearly justified for scalars of the first two generations, where
approximate degeneracy is predicted by the theory, and we make the
mild assumption that the third generation scalar masses do not differ
wildly from the first two.  Then, for any given scalar mass-squared
matrix, let us define an average diagonal entry by $\tilde{m}^2$, and
parameterize the size of the off-diagonal elements by
\begin{equation}
(\delta_{ij})^{\mbox{ }}_{AB}\equiv\frac{(\tilde{m}^2_{ij})^{\mbox{
}}_{AB}}{\tilde{m}^2} \,\,\, ,
\end{equation}
where $A$ and $B$ indicate chirality ($L$ or $R$), and the other
indices run over generation space.  Note that we evaluate the
$\delta_{ij}$ in the basis where the quark and lepton mass matrices
are diagonal.  This parameterization is especially useful since the
relevant experimental bounds on the $\delta_{ij}$ have been compiled
systematically by Gabbiani {\it et al.}\cite{mas}, allowing one to put
our model to the test.  In Tables~\ref{phone} through \ref{phsix}, we
give a comparison between the predictions of our model and the
experimental bounds.  The predictions follow directly from $\lambda$
power counting with $\lambda\approx 0.22$, and do not take into
account the undetermined order one coefficients.  Thus, the
predictions of our model may in fact be significantly smaller.
Table~\ref{phone} follows from $K$-$\overline{K}$, $D$-$\overline{D}$,
and $B$-$\overline{B}$ mixing constraints, Table~\ref{phtwo} from the
decay $b\rightarrow s \gamma$, and Table~\ref{phthree} from the lepton
flavor-violating decays $\ell_i\rightarrow \ell_j \gamma$.  Note that
the left-right scalar mass terms are determined by a Higgs VEV, as
well as the value of a dimensionful coefficient $A$.  The ratio of
Higgs VEV's is determined by the value of the bottom quark Yukawa
coupling that is predicted in our model; again, we take $\tan \beta
\approx 6$ for our estimates.  The $A$ parameter is of the same order
as the supersymmetry-breaking scale, but need not be equal to the
average scalar superpartner mass.  We give our assumed values for $A$
and the relevant superparticle masses in the table captions.  Of
course, taking larger values for the superparticle masses makes the
constraints less severe.

        To account for standard model CP violation, one must allow for
the possibility of complex operator coefficients and flavon VEV's.
Here one suffers from some ambiguity since the origins of CP violation
are not specified in our theory.  Nevertheless, one may compare the
constraints from CP-violating processes to the predictions of our
model assuming the worst-case scenario in which order one phases
appear in {\em all} entries of the scalar mass matrices.  The
corresponding results are given in Tables~\ref{phfour} through
\ref{phsix}.

        One learns from these tables that the flavor symmetry provides
a significant suppression of CP-conserving and CP-violating
flavor-changing processes, so that it is not necessary to take any
soft supersymmetry-breaking mass larger than $600$~GeV. It is
reasonable to view this as a positive result, especially when taking
into account the possible alternatives: For example, if the $(12)$
elements of all the $\tilde{m}^2$ matrices were of order the Cabibbo
angle $\lambda$, one would be forced to take the average squark mass
above $40$~TeV to evade the $K$-$\overline{K}$ mixing constraints, and
above $500$~TeV to evade the bounds on $\epsilon$, assuming order one
phases. The $D_6^L \times D_6^R$ symmetry is precisely what renders
the model suitable for solving the hierarchy problem, by allowing a
sufficiently low superparticle mass scale.  In addition, we see
that our results are compatible with the bounds from the
flavor-conserving, CP-violating electric dipole moments.  These bounds
could be accommodated even in the worst-case scenario of ubiquitous
order one phases, provided that $A$ terms are taken somewhat smaller
than the average scalar masses.

        In addition, these tables give a clear indication of where the
model is most likely to be tested experimentally.  The bounds that are
most marginally satisfied for light superparticle masses are those
from $\mu \rightarrow e \gamma$ and the various CP-violating
observables.  Improved bounds on $\mu \rightarrow e \gamma$ combined
with the discovery of light sleptons would provide the clearest means
of excluding the model.  In addition, improved bounds on electric
dipole moments, as well as a measurement of CP violation in the B
system, might exclude the generic order one phase picture described
above.  Depending on the precise bounds, the model may still be viable
in such a circumstance if CP is spontaneously broken in only some of
the flavon VEV's, since the pattern of phases in the soft masses would
then be far from generic.  However, we do not consider that
possibility further in this article.

\section{Conclusions}

        In this paper, we have explored the possibility of
constructing a realistic theory of flavor based on a relatively small
discrete gauge symmetry group.  Our $D^L_6 \times D^R_6$ model can
explain the hierarchy of quark and lepton masses via a three-step
sequential breaking pattern, and greatly alleviates the
flavor-changing neutral current problem that is endemic to generic,
softly-broken supersymmetric theories.  In addition, the smallness of
the flavor symmetry group prevents one from suppressing
flavor-changing effects arbitrarily, so that the model remains
falsifiable.  In particular, improved bounds on $\mu \rightarrow e
\gamma$ could be used to exclude the model if light sleptons are
eventually discovered. Better measurements of electric dipole moments
of the electron and neutron may demand modifying our assumption of a
generic pattern of order one phases in operator coefficients and
flavon VEV's, which we argued could account for standard model CP
violation, at least for some range of the soft supersymmetry-breaking
masses and $A$ parameters.  While in a small number of places we
needed to assume some fluctuation in order one coefficients, for
example to obtain the famous Georgi-Jarskog factors of $3$ in the
charged lepton Yukawa matrix, we do not view this as a serious
shortcoming of the model.  After all, we know of no low-energy
effective theory in particle physics where symmetry considerations
alone completely account for the particle phenomenology.  From this
point of view, our model works surprisingly well overall.  Finally,
there are a number of possible issues in our model that may be worth
additional investigation, such as spontaneous CP violation and
extension of the model to the neutrino sector.  However, it seems to
us that the more interesting problem is to find the most compelling
model based on a minimal discrete gauged flavor-symmetry group.  While
we believe we have made progress in the current work, the possibility
of constructing the most convincing theory of this type remains an
intriguing open question.

{\samepage
\begin{center}
{\bf Acknowledgments}
\end{center}
CC thanks the National Science Foundation for support under Grant No.\
PHY-9800741; RL thanks the Department of Energy for support under
Contract No.\ DE-AC05-84ER40150.}

\appendix

\section{Group Theory Fundamentals} \label{grpthy}

        In this appendix we describe in general terms the group theory
necessary to understand the manipulation of discrete horizontal
symmetry groups, and re-acquaint the reader with the necessary
terminology.  The fundamentals of discrete group theory used here
appear, for example, in the book by Hamermesh\cite{Ham}, particularly
the first four chapters.

        An abstract {\it group\/} $G$ is defined as any set of objects
closed under an associative binary operation, called multiplication;
the set contains a multiplicative identity; and each element has a
multiplicative inverse.  Clearly, the structure of $G$ is completely
determined by a multiplication table of its elements.  Two groups are
{\it isomorphic\/} if they have the same number of elements and the
same multiplication table.

	A {\it representation\/} $R$ of $G$ is just a set of objects
satisfying the same multiplication table as $G$, although it can
happen that two or more distinct elements of $G$ correspond to only
one element of $R$ (that is, $R$ considered as a group need not be
isomorphic to $G$); moreover, the actual operation corresponding to
multiplication in $R$ may be very different from that used by $G$.
Operationally, the most useful group representations are in terms of
$n \times n$ matrices under the usual matrix multiplication ($n$-{\it
dimensional} or $n$-{\it plet representations}), where some of
the representation matrices may fail to commute, providing a natural
way to accommodate non-Abelian groups.  A representation is {\it
irreducible\/} if the corresponding matrices cannot be simultaneously
block-diagonalized, and two matrix representations are {\it
equivalent\/} if they can be related by a change of basis.

        The most interesting feature of finite groups for our purposes
is that the number of elements ({\it order\/}) $g$ of $G$ is related
to the number and dimensions $n_\nu$ of its inequivalent irreducible
representations $\nu$,
\begin{equation} \label{repcount}
g = \sum_\nu n_\nu^2 ,
\end{equation}
meaning that a given finite group has only a finite number of
inequivalent irreducible representations, all of them
finite-dimensional.  It can be shown that all irreducible
representations of an Abelian group are one-dimensional.

	Complete information on the representation content of a given
group is exhibited by its {\it character table}.  The {\it
character\/} of an element of $R$ is just the trace of the
corresponding matrix, which is invariant under basis changes.
Reference \cite{Ham} presents character tables for many common small
groups, while Ref.~\cite{GrT} presents full information for all groups
of order $\le 31$.

\section{Manipulation of $D_6$ Representations} \label{clebsch}

        The group $D_6$ has four inequivalent singlet representations,
${\bf 1_S^+}$, ${\bf 1_A^+}$, ${\bf 1_S^-}$, and ${\bf 1_A^-}$, and
two inequivalent doublet representations, ${\bf 2^-}$ and ${\bf 2^+}$.
In the notation of Ref.~\cite{FKep}, these are $1$, $1^\prime$,
$1^{\prime\prime}$, $1^{\prime\prime\prime}$, $2_{(1)}$, and
$2_{(2)}$, respectively.

        The singlet representations of $D_6$ obey the same
multiplication rules as elements of $Z_2 \times Z_2$, where the
even/odd elements of the first $Z_2$ are {\bf S}/{\bf A}, and those of
the second are $+/-$.  Thus, for example, ${\bf 1_A^-} \otimes {\bf
1_S^-} = {\bf 1_A^+}$.  The full list of binary tensor products of
singlet representations appears in Table~\ref{onexone}.

        For doublet (and larger) representations, the question of
combining two representations is less trivial, and constitutes the
question of computing Clebsch-Gordan coefficients of a given group, in
this case $D_6$.  The general expression relating the product of an
$n_x$-plet ${\bf x}$ and and an $n_y$-plet ${\bf y}$ to an $n_z$-plet
${\bf z}$ is given by
\begin{equation}
R_x^\dagger {\cal O}_i R_y = \sum_{j=1}^{n_z} (R_z)_{ij} {\cal O}_j,
\hspace{2em} i=1, \ldots , n_z,
\end{equation}
where $R$ refers to each matrix in a given representation of $D_6$,
and the $n_z$ matrices ${\cal O}_i$ of dimensions $n_x \times n_y$ are
the Clebsch-Gordan coefficients of the group.  That is, ${\bf
x}^\dagger {\cal O}_i {\bf y}$ transforms like the $i$th component of
${\bf z}$.  Since the dihedral groups $D_n$ are entirely generated by
the rotations $C_n$ and $C_x$, it is enough to obtain ${\cal O}$
satisfying this condition for just $R = C_n, C_x$.  Explicit
representations of the $C_x$, $C_6$ are given in Table~\ref{reps}.
The tensor products of singlet with doublet representations are given
by
\begin{eqnarray} \label{onextwo}
{\bf 1}_{\bf S}^{P_1} \otimes {\bf 2}^{P_2} & = & {\bf 2}^{P_1 P_2},
\mbox{ with } {\cal O}_1 = (1 \,\,\, 0 ), \,\,\, {\cal O}_2 = (
\,\,\,\, 0 \,\,\, 1), \nonumber \\
{\bf 1}_{\bf A}^{P_1} \otimes {\bf 2}^{P_2} & = & {\bf 2}^{P_1 P_2},
\mbox{ with } {\cal O}_1 = (0 \,\,\, 1 ), \,\,\, {\cal O}_2 = (-1
\,\,\, 0),
\end{eqnarray}
while the products of two doublet representations are given by
\begin{equation}
{\bf 2}^{P_1} \otimes {\bf 2}^{P_2} = {\bf 2}^{P_1 P_2} \oplus {\bf
1}_{\bf A}^{P_1 P_2} \oplus {\bf 1}_{\bf S}^{P_1 P_2} ,
\end{equation}
where the Clebsch-Gordan coefficients are given by
\begin{eqnarray}
{\bf 2}^{P_1} \otimes {\bf 2}^{P_2} & \supset & {\bf 2}^{P_1 P_2}:
{\cal O}_1 = \sigma_3, \, {\cal O}_2 = -\sigma_1; \nonumber \\ &
\supset & {\bf 1}_{\bf A}^{P_1 P_2}: \, {\cal O} = i\sigma_2;
\nonumber \\ & \supset & {\bf 1}_{\bf S}^{P_1 P_2}: \, {\cal O} =
\openone.
\end{eqnarray}
Here $\sigma_i$ are the standard Pauli matrices, and $\openone$ is the
identity matrix.  That the parities $P_i$ are trivially multiplicative
when representations are combined is clear from the isomorphism $D_6
\cong D_3 \times Z_2$.  Note also from Eq.~(\ref{onextwo}) that the
effect of forming the tensor product between either ${\bf 2}$ with
${\bf 1_A^{\pm}}$ has the effect of switching the order of the
components in the doublet, while that with ${\bf 1_S^{\pm}}$ leaves
the order unchanged.  We refer to a doublet whose components have been
interchanged in this way generically as a ${\bf \tilde 2}$.

\begin{table}

\begin{tabular}{l|cc|cc}

Flavon receiving VEV & Magnitude of VEV && Remaining symmetry & \\
\hline

$\sigma_b$, $\sigma_\tau$, $\phi^{(2)}$, $\psi^{(2)}$,
$\Sigma^{(2,2)}$ & $\lambda^2$ && $Z_2^L \times Z_2^R$ & \\

$\phi^{(1)}$, $\Sigma^{(1,2)}$ & $\lambda^3$ && $Z_1^L \times Z_2^R$ &
\\

$\psi^{(1)}$, $\Sigma^{(1,1)}$, $\Sigma^{(2,1)}$ & $\lambda^4$ &&
$Z_1^L \times Z_1^R$ &

\end{tabular}

\caption{Pattern of sequential symmetry breaking in $D_6^L \times
D_6^R$.  $Z_2$ factors refer to the twofold symmetry of the $D_3$
subgroup and not the parity subgroup of $D_6$, while $Z_1$ is
shorthand for no symmetry.  Superscripts refer to particular tensor
components, which receive separate VEV's as described in the text.}
\label{sym}

\end{table}

\begin{table}
\begin{center}
\begin{tabular}{cccc}
& $\sqrt{|{\rm Re}(\delta_{12}^d)^{2}_{LL}|}$ & $\sqrt{|{\rm
Re}(\delta_{12}^d)^{2}_{LR}|}$ & $\sqrt{|{\rm
Re}(\delta_{12}^d)^{\mbox{ }}_{LL}(\delta_{12}^d)^{\mbox{ }}_{RR}|}$
\\ \hline Model & $5.2\times 10^{-4}$ & $4.0\times 10^{-5}$ & $2.4
\times 10^{-4}$ \\
Expt.  &  $4.0\times 10^{-2}$ & $4.4\times 10^{-3}$  & $2.8 \times
10^{-3}$ \\
\hline & $\sqrt{|{\rm Re}(\delta_{13}^d)^{2}_{LL}|}$
&$\sqrt{|{\rm Re}(\delta_{13}^d)^{2}_{LR}|}$ & $\sqrt{|{\rm
Re}(\delta_{13}^d)^{\mbox{ }}_{LL}(\delta_{13}^d)^{\mbox{ }}_{RR}|}$
\\ \hline Model & $1.1\times 10^{-2}$ & $4.0\times 10^{-5}$ &
$5.0\times 10^{-3}$ \\
Expt.  & $9.8\times 10^{-2}$ & $3.3\times 10^{-2}$ & $1.8\times
10^{-2}$
\\ \hline & $\sqrt{|{\rm Re}(\delta_{12}^u)^{2}_{LL}|}$ & $\sqrt{|{\rm
Re}(\delta_{12}^u)^{2}_{LR}|}$ &
$\sqrt{|{\rm Re}(\delta_{12}^u)^{\mbox{ }}_{LL}(\delta_{12}^u)^{\mbox{
}}_{RR}|}$ \\ \hline Model & $5.2\times 10^{-4}$ &$2.5\times 10^{-4}$
&$2.4\times 10^{-4}$ \\ Expt.  & $1.0\times 10^{-1}$ &$3.1\times
10^{-2}$ &$1.7\times 10^{-2}$
\end{tabular}
\end{center}
\caption{Bounds on the off-diagonal squark mixing parameters
$\delta_{ij}$ from neutral meson mixing, assuming a gluino and average
squark mass of $500$~GeV. For left-right masses, an $A$ parameter of
$500$~GeV was also assumed. Note that the experimental bounds scale
linearly in the average squark mass, given a fixed ratio with the
gluino mass. \label{phone}}
\end{table}

\begin{table}
\begin{center}
\begin{tabular}{ccc}
& $|(\delta_{23}^d)^{\mbox{ }}_{LL}|$ & $ |(\delta_{23}^d)^{\mbox{
}}_{LR}|$ \\ \hline Model & $4.8\times 10^{-2}$ & $1.8\times 10^{-4}$
\\ Expt.  & $8.2$ & $1.6\times 10^{-2}$
\end{tabular}
\end{center}
\caption{Bounds on $\delta_{ij}$ from $b\rightarrow s \gamma$,
assuming a gluino and average squark mass of $500$~GeV. For left-right
masses, an $A$ parameter of $500$~GeV was also assumed. Note that the
experimental bounds scale quadratically in the average squark
mass, given a fixed ratio with the
gluino mass.\label{phtwo}}
\end{table}

\begin{table}
\begin{center}
\begin{tabular}{ccc}
& $|(\delta_{12}^\ell)^{\mbox{ }}_{LL}|$ & $
|(\delta_{12}^\ell)^{\mbox{ }}_{LR}|$ \\ \hline Model & $5.2\times
10^{-4}$ & $1.6\times 10^{-5}$ \\
Expt.  & $9.4\times 10^{-2}$ & $2.1\times 10^{-5}$ \\ \hline &
$|(\delta_{13}^\ell)^{\mbox{ }}_{LL}|$ & $ |(\delta_{13}^\ell)^{\mbox{
}}_{LR}|$
\\ \hline Model & $1.1\times 10^{-2}$ & $1.6\times 10^{-5}$ \\ Expt.
& $355$ & $1.3$ \\ \hline
& $|(\delta_{23}^\ell)^{\mbox{ }}_{LL}|$ & $
|(\delta_{23}^\ell)^{\mbox{ }}_{LR}|$ \\ \hline
Model &  $4.8\times 10^{-2}$ & $7.3\times 10^{-5}$ \\
Expt.  &  $65$ & $2.5\times 10^{-1}$
\end{tabular}
\end{center}
\caption{Bounds on the $\delta_{ij}$ from $\ell_i \rightarrow \ell_j
\gamma$ decays, assuming a photino and average slepton mass of
$350$~GeV. For left-right masses, an $A$ parameter of $100$~GeV was
also assumed. Note that the experimental bounds scale quadratically in
the
average slepton mass, given a fixed ratio with the photino mass. 
\label{phthree}}
\end{table}

\begin{table}
\begin{center}
\begin{tabular}{cccc}
& $\sqrt{|{\rm Im}(\delta_{12}^d)^2_{LL}|}$ &
$\sqrt{|{\rm Im}(\delta_{12}^d)^2_{LR}|}$ &
$\sqrt{|{\rm Im}(\delta_{12}^d)^{\mbox{ }}_{LL}(\delta_{12}^d)^{\mbox{
}}_{RR}|}$ \\ \hline
Model & $5.2\times 10^{-4}$ & $3.3\times 10^{-5}$ & $2.4 \times
10^{-4}$ \\
Expt.  &  $3.8\times 10^{-3}$ & $4.2\times 10^{-4}$ & $2.6\times
10^{-4}$
\end{tabular}
\end{center}
\caption{Bounds on the $\delta_{ij}$ from $\epsilon$ assuming a gluino
and average squark mass of $600$~GeV. For left-right masses, an $A$
parameter of $600$~GeV was also assumed. Note that the experimental
bounds scale linearly in the average squark mass, given a fixed ratio 
with the gluino mass. The model predictions assume a worst-case 
scenario in which order one phases appear in all entries of the scalar 
mass matrices. \label{phfour}}
\end{table}

\begin{table}
\begin{center}
\begin{tabular}{ccc}
& $|{\rm Im}(\delta_{12}^d)^{\mbox{ }}_{LL}|$ &
$|{\rm Im}(\delta_{12}^d)^{\mbox{ }}_{LR}|$ \\ \hline
Model &  $5.2\times 10^{-4}$ & $1.6\times 10^{-5}$ \\
Expt.  &  $6.9\times 10^{-1}$ & $2.9\times 10^{-5}$
\end{tabular}
\end{center}
\caption{Bounds on the $\delta_{ij}$ from $\epsilon'/\epsilon$,
assuming a gluino and average squark mass of $600$~GeV. For left-right
masses, an $A$ parameter of $300$~GeV was also assumed. Note that the
experimental bounds scale quadratically in the average squark
mass, given a fixed ratio with the gluino mass. The model predictions 
assume a worst-case scenario in which order one phases appear 
in all entries of the scalar mass matrices.
\label{phfive}}
\end{table}

\begin{table}
\begin{center}
\begin{tabular}{cccc}
& $|{\rm Im}(\delta_{11}^d)^{\mbox{ }}_{LR}|$ & $|{\rm
Im}(\delta_{11}^u)^{\mbox{ }}_{LR}|$ & $|{\rm
Im}(\delta_{11}^\ell)^{\mbox{ }}_{LR}|$ \\ \hline
Model & $3.0\times 10^{-6}$ & $4.2\times 10^{-6}$ & $1.8\times
10^{-6}$
\\
Expt.  & $3.6\times 10^{-6}$ & $7.1\times 10^{-6}$ & $2.2\times
10^{-6}$
\end{tabular}
\end{center}
\caption{Bounds on the $\delta_{ii}$ from electric dipole moments,
assuming a gluino, photino, average squark, and average slepton mass of
$600$~GeV.  An $A$ parameter of $250$~GeV for the squarks and
$150$~GeV for the sleptons was assumed. The model predictions assume a
worst-case scenario in which order one phases appear in all entries of
the scalar mass matrices.
\label{phsix}}
\end{table}

\begin{table}

\begin{tabular}{ccc||cc|cc|cc|cc}

& $\otimes$ && ${\bf 1_S^+}$ && ${\bf 1_A^+}$ && ${\bf 1_S^-}$ &&
${\bf 1_A^-}$ & \\ \hline

& ${\bf 1_S^+}$ && ${\bf 1_S^+}$ && ${\bf 1_A^+}$ && ${\bf 1_S^-}$ &&
${\bf 1_A^-}$ & \\

& ${\bf 1_A^+}$ && ${\bf 1_A^+}$ && ${\bf 1_S^+}$ && ${\bf 1_A^-}$ &&
${\bf 1_S^-}$ & \\

& ${\bf 1_S^-}$ && ${\bf 1_S^-}$ && ${\bf 1_A^-}$ && ${\bf 1_S^+}$ &&
${\bf 1_A^+}$ & \\

& ${\bf 1_A^-}$ && ${\bf 1_A^-}$ && ${\bf 1_S^-}$ && ${\bf 1_A^+}$ &&
${\bf 1_S^+}$ &

\end{tabular}

\caption{Tensor products of two singlet representations of
$D_6$.} \label{onexone}

\end{table}

\begin{table}

\begin{tabular}{ccc||cc|cc|cc|cc|cc|cc}

&&& ${\bf 1_S^+}$ && ${\bf 1_A^+}$ && ${\bf 1_S^-}$ && ${\bf 1_A^-}$
&& ${\bf 2^+}$ && ${\bf 2^-}$ & \\ \hline

& $C_x$ && $+1$ && $-1$ && $+1$ && $-1$ && $\left( \begin{array}{cc} 1
& 0 \\ 0 & -1 \end{array} \right)$ && $\left( \begin{array}{cc} 1 & 0
\\ 0 & -1 \end{array} \right)$ \\ \hline

& $C_6$ && $+1$ && $+1$ && $-1$ && $-1$ && $\left( \begin{array}{cc}
-\frac 1 2 & +\frac{\sqrt{3}}{2} \\ -\frac{\sqrt{3}}{2} & -\frac 1 2
\end{array} \right)$ && $\left( \begin{array}{cc}
+\frac 1 2 & -\frac{\sqrt{3}}{2} \\ +\frac{\sqrt{3}}{2} & +\frac 1 2
\end{array} \right)$

\end{tabular}

\caption{Explicit representations of the generating elements of $D_6$
for each inequivalent irreducible representation.} \label{reps}

\end{table}

\end{document}